\documentstyle[aps,prl,multicol,epsfig]{revtex}

\begin{document}

\newcommand{\beq}{\begin{eqnarray}}
\newcommand{\eeq}{\end{eqnarray}}
\renewcommand{\thefootnote}{\fnsymbol{footnote}}

\title{Nonequilibrium probabilistic dynamics of the logistic  map at the edge of chaos}

\author{Ernesto P. Borges$^{1,2}$, 
Constantino Tsallis$^{1}$, 
Gar\'\i n F. J. A\~na\~nos$^{1,3}$, 
Paulo Murilo C. de Oliveira$^{4}$
}
\address{
$^1$Centro Brasileiro de Pesquisas F\'\i sicas, \\
R. Dr.  Xavier Sigaud 150, 
22290-180 Rio de Janeiro, RJ, Brazil \\
$^2$Escola Polit\'ecnica, 
Universidade Federal da Bahia, \\
R. Aristides Novis, 2, 40210-630 Salvador, BA, Brazil \\
$^3$Departamento de F\'{\i}sica, Universidad Nacional de Trujillo \\
Av. Juan Pablo II, s/n, Trujillo, Peru \\
$^4$Instituto de F\'\i sica, 
Universidade Federal Fluminense, \\
Av. Litor\^anea s/n, Boa Viagem, 24210-340, Niter\'oi, RJ, Brazil
}
\maketitle
\begin{abstract}
We consider nonequilibrium probabilistic dynamics in logistic-like maps 
$x_{t+1}=1-a|x_t|^z$, $(z>1)$ at their chaos threshold: We first introduce many
initial conditions within one among $W>>1$ intervals partitioning the phase 
space and focus on the unique value $q_{sen}<1$ for which the entropic form 
$S_q \equiv \frac{1-\sum_{i=1}^{W} p_i^q}{q-1}$
{\it linearly} increases with time. We then verify that 
$S_{q_{sen}}(t) - S_{q_{sen}}(\infty)$  vanishes like 
$t^{-1/[q_{rel}(W)-1]}$ [$q_{rel}(W)>1$]. We finally exhibit a new finite-size 
scaling, $q_{rel}(\infty) - q_{rel}(W) \propto W^{-|q_{sen}|}$. 
This establishes quantitatively, for the first time, a long pursued relation
between sensitivity to the initial conditions and relaxation, 
concepts which play central roles in nonextensive statistical mechanics.

$ $

\end{abstract}


\begin{multicols}{2}

  Connections between dynamics and thermodynamics are not at all 
  completely elucidated. 
  Frequently thermostatistics may sound as if it would consist only of a 
  self-referred body, which could dispense dynamics from its formulation.
  But this has long been known not to be true  (see \cite{egdcohen} and
  references therein).
  One possible reason for this essential point having been poorly emphasized is 
  that when dealing with weakly interacting systems, Boltzmann-Gibbs (BG)
  thermodynamical equilibrium may be formulated without referring to the 
  underlying dynamics of its constituents.
  As complex systems came to the front line of research
  (by complex we mean here the presence of at least one of the features:
  long-range interparticle interactions, long-term memory, fractal nature of 
  a pertinent phase-space, small-world networking),
  it became necessary to revisit this fundamental issue \cite{baranger}.
  Indeed, a significant number of systems, e.g., 
turbulent fluids \cite{turbulentbeck,turbulentarimitsu}, 
electron-positron annihilation \cite{bediaga}, 
economics \cite{borland}, 
motion of {\it Hydra viridissima} \cite{hydra}, 
and others, are known nowadays to be not properly described with 
BG statistical mechanical concepts.
Systems such as these have been successfully handled within 
nonextensive statistical mechanics.
  The purpose of this Letter is to numerically illustrate, 
  with a very simple dynamical model (logisticlike maps), that basic dynamical 
  concepts such as the sensitivity to the initial conditions and the relaxation
  towards equilibrium are deeply entangled, thus yielding a new analytic 
  scaling connection.

  The basic building block of nonextensive statistical mechanics is 
  the nonextensive entropy \cite{CT1988}
  \beq
  S_q \equiv \frac{1-\sum_{i=1}^{W} p_i^q}{q-1} \qquad (q \in {\cal R}).
  \label{sq}
  \eeq
  The entropic index $q$ characterizes the statistics we are dealing with; 
  $q=1$ recovers the usual BG expression, $S_1=-\sum_{i=1}^W p_i \ln p_i$.
  We may think of $q$ as a biasing parameter: $q<1$ privileges rare events,
  while $q>1$ privileges ordinary events:
  $p<1$ raised to a power $q<1$ yields a value greater than $p$, 
  and the relative increase $p^q/p=p^{q-1}$ is a {\it decreasing}
  function of $p$, i.e., values of $p$ closer to 0 (rare events) are benefited.
  Correspondingly, for $q>1$, values of $p$ closer to 1 (ordinary events) are 
  privileged.
  BG ($q=1$) is the unbiased statistics.
  A concrete consequence of this is that the BG formalism yields exponential 
  equilibrium distributions,
  whereas nonextensive statistics yields power-law distributions 
  (the BG exponential is recovered as a limiting case: 
  we are talking of a {\it generalization}, not an alternative).
Many developments concerning this formalism are available in the literature,
\cite{Curado_Tsallis,CTBJP},
but there are still important points that remain to be better understood.
One of them is what lays behind the quite impressive results recently 
put forward for fully developed turbulence.
Agreement between theoretical treatments and experimental data has been 
achieved along two different approaches, both based on nonextensive
statistical mechanics, but, at first sight surprisingly, one of them with $q>1$ \cite{turbulentbeck}
and the  other with $q<1$ \cite{turbulentarimitsu}.
A similar scenario (using both $q<1$ and $q>1$) is also found in the 
description of low dimensional dynamical systems, particularly in the
$z$-logistic maps at the threshold of chaos.
This Letter intends to shed light on this problem, by investigating these
maps both in the chaotic region and at the edge of chaos, and by showing, 
for the first time, a long pursued connection, namely, 
between the {\it sensitivity}-based and the {\it relaxation}-based approaches.

Consider the $z$-logistic iterative equation
%
$x_{t+1}=1-a|x_t|^z$ 
%
$(-1\le x_t \le 1;\; 0 < a \le 2;\; z>1;\; t=0,1,2 \dots)$
for which $z=2$ recovers the logistic map,
and partition the phase space into $W$ cells of equal measure, 
$p_i$ being the probability to be soon associated with the $i^{th}$ cell.
Several works \cite{Costa-Lyra-Plastino-CT,Lyra-CT,Tirnakli-Tsallis-Lyra,Latora-Baranger-Rapisarda-CT,Tirnakli-Ananos-CT}
have shown that the description of such dynamical systems is properly achieved
by using the nonextensive entropy, Eq.~(\ref{sq}),
with a specific $z$-dependent value of $q$.

Let us first recall the sensitivity-based approaches, which determine 
$q_{sen} \le 1$ ({\it sen} stands for sensitivity).
Up to now, three different methods have been developed which provide the 
special value $q_{sen}$ (sometimes noted as $q^*$). 
The first method \cite{Costa-Lyra-Plastino-CT}
is based on the sensitivity to initial conditions. At the threshold of
chaos $[a=a_c(z)]$, the Lyapunov exponent $\lambda_1$ vanishes 
and the system is weakly sensitive ({\it weak} chaos), i.e., the upper bound 
separation between nearby trajectories 
$\xi(t) \equiv \lim_{\Delta x(0) \to 0} \frac{\Delta x(t)}{\Delta x(0)}$ 
typically evolves \cite{baldovinrobledo} in time according to the power-law
\beq
\label{sensitivity}
\xi(t) = [1+(1-q_{sen})\lambda_{q_{sen}} \, t]^{\frac{1}{1-q_{sen}}}
\qquad (\lambda_{q_{sen}} > 0),
\eeq
which is solution of the nonlinear differential equation 
$\dot{\xi}=\lambda_{q_{sen}} \, \xi^{q_{sen}}$ (this equation recovers, 
for $q_{sen}=1$,  the usual exponential sensitivity to the initial conditions, 
referred to as {\it strong} chaos).
The second method is based on the geometrical description of the multifractal
attractor \cite{Lyra-CT,Tirnakli-Tsallis-Lyra}.
The value of $q_{sen}$ is determined by  
%
$\frac{1}{1-q_{sen}}=\frac{1}{\alpha_{min}}-\frac{1}{\alpha_{max}}$, 
%
where $\alpha_{min}$ and $\alpha_{max}$ are the values of the end points
of the multifractal function $f(\alpha)$.
This beautiful equation relates dynamics (left-hand side) with geometry 
(right-hand side).
The third method is related, as we shall specify, to a conjectured 
generalization of the Pesin relation 
\cite{Latora-Baranger-Rapisarda-CT,Tirnakli-Ananos-CT}. 
In one of the $W$ cells in which the phase space has been divided, we 
initially put (randomly or uniformly distributed) $N$ points  $(N \gg W)$. 
We then follow the occupancies $\{N_i(t)\}$ of all cells 
[$\sum_{i=1}^W N_i(t)=N$], which enable the calculation of the probabilities 
$p_i(t) \equiv N_i(t)/N$, which enable in turn the calculation of 
$S_q(t)$ as given by Eq.~(\ref{sq}). We can next define a nonextensive 
version of the Kolmogorov-Sinai entropy, namely
$K_{q} \equiv \lim_{t\rightarrow \infty} \lim_{W\rightarrow \infty}
\lim_{N\rightarrow \infty} S_{q}(t)/t$. The special value $q=q_{sen}$ is the 
one for which the entropy production $K_{q}$ is {\it finite} 
(if $q<q_{sen}$, $K_q \rightarrow \infty$, and if $q>q_{sen}$, 
$K_q \rightarrow 0$). 
It is quite remarkable that all three methods give (within an acceptable 
numerical error) one and the same value for $q_{sen}$  ($q_{sen}=0.2445...$ for $z=2$).

Let us now turn to the relaxation-based approach of the problem, 
first tackled in \cite{Moura-Tirnakli-Lyra}. We start now 
\cite{Moura-Tirnakli-Lyra} with an ensemble of initial conditions uniformly
spread over the entire phase space (a procedure that resembles the Gibbsian
approach of statistical mechanics), instead of within a single among the 
$W$ cells (closer to a Boltzmannian approach), and investigate the rate of 
convergence onto the multifractal attractor at the onset of chaos.
At $t=0$ all $W$ cells are occupied (hence $S_q= \frac{W^{1-q}-1}{1-q}$). 
As $t$ increases, the number $W_{occupied}$ of cells that have at least 
one point inside typically decreases (shrinking of the Lebesgue measure) as 
$1/[1+(q_{rel}-1)\;t/\tau_q]^{1/(q_{rel}-1)}$ 
({\it rel} stands for relaxation; $q_{rel} \ge 1$; $\tau_q>0$ is a 
characteristic relaxation time).

When applied to the chaos region (e.g., $z=2$, $a=2$),
all these methods yield $q_{sen}=q_{rel}=1$, in accordance with the
usual BG concepts, founded on strongly chaotic systems,
consistently with Boltzmann's ``molecular chaos'' hypothesis (see \cite{Ademir} 
and references therein). When applied to the edge of chaos $a_c(z)$ of the 
$z$-logistic maps, it is obtained \cite{Moura-Tirnakli-Lyra} $q_{rel}(\infty)$ 
[$q_{rel}(\infty)=2.4...$ for $z=2$].
In this Letter, we start with $N$ points (we adopt $N=10\,W$ for 
numerical convenience) uniformly distributed inside a {\it single} 
specific cell chosen as soon specified. We compute $S_{q_{sen}}(t)$ and 
verify that, for a fixed value of $W$, $S_{q_{sen}}(t)$ asymptotically reaches
$S_{q_{sen}}(\infty)$ [$S_{q_{sen}}(\infty)>0$  monotonically increases 
with $W$].  $S_{q_{sen}}(t)$ reaches its $t \to \infty$ value from above, 
preceded by an overshooting, which might be very strong. 
The initial cell is chosen as that one which presents the highest overshooting.

If we consider the strongly chaotic case $a=2$, hence $q_{sen}=1$,
$S_1(t)$ slightly overshoots and rapidly saturates, as illustrated in Fig.~1(a) for $z=2$.
At the onset of chaos, however, $S_{q_{sen}}(t)$ presents a very pronounced
overshoot, and slowly approaches its final value $S_{q_{sen}}(\infty)$ 
[Fig.~1(b)]. 
We then compute, for all times after the overshooting, 
$\Delta S_{q_{sen}}(t) \equiv [S_{q_{sen}}(t)-S_{q_{sen}}(\infty)]$.
We find a power-law decay, 
whose log-log slope  depends on $W$.
(Log-periodic oscillations such as those reported in \cite{Moura-Tirnakli-Lyra}
are also detected here, and are better visualized 
for $z<2$.). 
We identify this slope with $1/(q_{rel}(W)-1)$ 
(as an alternative way of obtaining $q_{rel}$).
In the limit $W \rightarrow \infty$, we observe that $q_{rel}(\infty)$ 
{\it precisely coincides}, for all values of $z$,  
with the  value found in Ref.\cite{Moura-Tirnakli-Lyra}
(their values correspond to infinitely fine graining $W \to \infty$).
The power-law region of $\Delta S_{q_{sen}}(t)$ increases with $W$ (see Fig.~2). 
In order to better identify this region, we consider only those time 
intervals whose linear correlation coefficient is larger than
0.99.  In order to increase the precision of $q_{rel}(W)$ we take averages
over various possible intervals, more precisely, ranging from various 
initial and various final times, as shown in Fig.~2. 
Moreover, in order to further minimize the effect [on the precision of 
$q_{rel}(W)$] of the oscillations of $S_{q_{sen}}(t)$ 
at the onset of chaos, and also to better identify the attractor 
--- the invariant distribution --- we introduce a numerical improvement in the 
iteration rule which does not affect the asymptotic dynamics of the map:
we average the distribution corresponding to  
$x_{t}$ with that corresponding to $x_{t-1}$ 
(the actual values of $x$ must be preserved while averaging).
This procedure strongly reduces the considerable fluctuations in the distribution of points at $t \rightarrow \infty$ 
and enables  us to identify the attractor associated with the map 
(we refer to the attractor in the space of distributions, 
and not in phase space).
The computer implementation of these numerical benefits demands additional 
memory and an increase of the CPU time. 
The oscillations that remain are considerably smaller, and they can be 
further reduced by extending the method in order to take into account more 
than one previous time step. However, for the accuracy we seek in this work, 
it is clearly enough to average the distributions of $x_{t-1}$ and $x_{t}$.
We emphasize that this kind of averaging is just a trick to minimize the
fluctuations (the original dynamics is preserved, once we average the
distribution of $x$, and not $x$ itself). If the procedure is not
adopted, we find the same results, though with less precision.
After all these numerical improvements have been implemented, a remarkable law 
emerges, namely,
\beq
\label{scaling}
q_{rel}(\infty) - q_{rel}(W) \propto W^{-|q_{sen}|}
\eeq
(see Fig.~3).
This equation is somehow analogous to the expressions that appear
in finite-size scaling \cite{Barber}.
This law is quite impressive, since it relates two basic dynamical properties 
of dissipative systems, namely, relaxation onto equilibrium and sensitivity 
to the initial conditions (the basis for mixing).
Moreover,  the (coarser or finer) graining --- $1/W$ for the $z$-logistic
maps --- is involved, pretty much in the same way as it occurs with the values
of $q$ (here noted as $q_{rel}$) appearing in Beck's approach 
\cite{turbulentbeck} of fully developed turbulence, where the entropic index 
depends on the distance between the two points at which the fluid velocities 
are measured.
Other phenomena where a similar scaling relation may occur are
the distribution of transverse momenta of hadronic jets produced in
electron-positron annihilation experiments \cite{bediaga},
and the saddle-point dynamics of the Henon-Heiles system \cite{Ivano}. 
Let us also mention that the scaling relation, Eq.~(\ref{scaling}), 
has a moderate 
sensitivity to the value used for $q_{sen}$ (see the Inset of Fig.~3).
It is important to emphasize that Eq.~(\ref{scaling})
reflects the fact that the limits 
$\lim_{t\rightarrow \infty}$ $\lim_{W\rightarrow \infty}$ 
(relevant for $q_{sen}$; see for instance \cite{Latora-Baranger-Rapisarda-CT}) and 
$\lim_{W\rightarrow \infty} \lim_{t\rightarrow \infty}$ 
(relevant for $q_{rel}$, as exhibited here) do {\it not} commute 
(nonuniform convergence) in general (in other words, generically 
$q_{sen} \ne q_{rel}$, whereas for strong chaos it is $q_{sen} =q_{rel}=1$). 
This result is similar to the noncommuting $t\rightarrow \infty$ and 
$M\rightarrow \infty$ limits for long-range interacting $M$-body Hamiltonian 
systems \cite{Latora-Rapisarda-CT}, for which vanishing Lyapunov exponents 
have already been detected \cite{Celia} (we remind that these limits do commute
for short-range interactions).
We finally exhibit Fig.~4, which shows $q_{rel}(\infty)$, as well as
$q_{sen}$ (taken from \cite{Costa-Lyra-Plastino-CT}), for typical values of $z$.
For $q_{rel}(\infty)$ we have indeed verified the coincidence
(with high precision for all $z$) with the values observed 
in \cite{Moura-Tirnakli-Lyra} by starting with uniform occupation of the 
phase space and following the shrinking of the Lebesgue measure. In addition 
to these results, we observe that
$q_{sen}(z \rightarrow \infty) \simeq 0.72$ and
$ \lim_{W \rightarrow \infty}q_{rel}(z \rightarrow 1) \simeq 4/3$.

  In conclusion, the scenario which emerges is that the thermal equilibrium of 
  all Hamiltonian systems whose dynamics is dictated by 
  noninteracting or short-range interacting particles  
  is described by Boltzmann-Gibbs statistical mechanics, 
  and usual thermodynamics apply, typically together with the usual exponential
  relaxation to equilibrium. 
  But for systems with complex dynamics (complex in the sense previously 
  described), we cannot know {\it a priori} how the relaxation to the 
  stationary state occurs. However, for a vast class of such systems, power-law
  relaxation does occur. To determine the associated exponent we must analyze 
  at least once the dynamics of that particular nonextensivity universality 
  class, in such a way as to determine once the corresponding values of $q$.
  For $z$-logistic maps, Eq.~(\ref{scaling}) (main result of the present work) 
  gives the connection between those values of $q$. The analogous task for 
  Hamiltonian systems would be very welcome.

\section*{Acknowledgments}
One of us (CT) thanks H.J. Herrmann for fruitful remarks.
This work is partially supported by PRONEX/MCT, CNPq, CAPES and FAPERJ
(Brazilian agencies).


\end{multicols}

\newpage
\begin{figure}[htb]
\begin{center}
\includegraphics[width=0.75\textwidth,keepaspectratio,clip=]{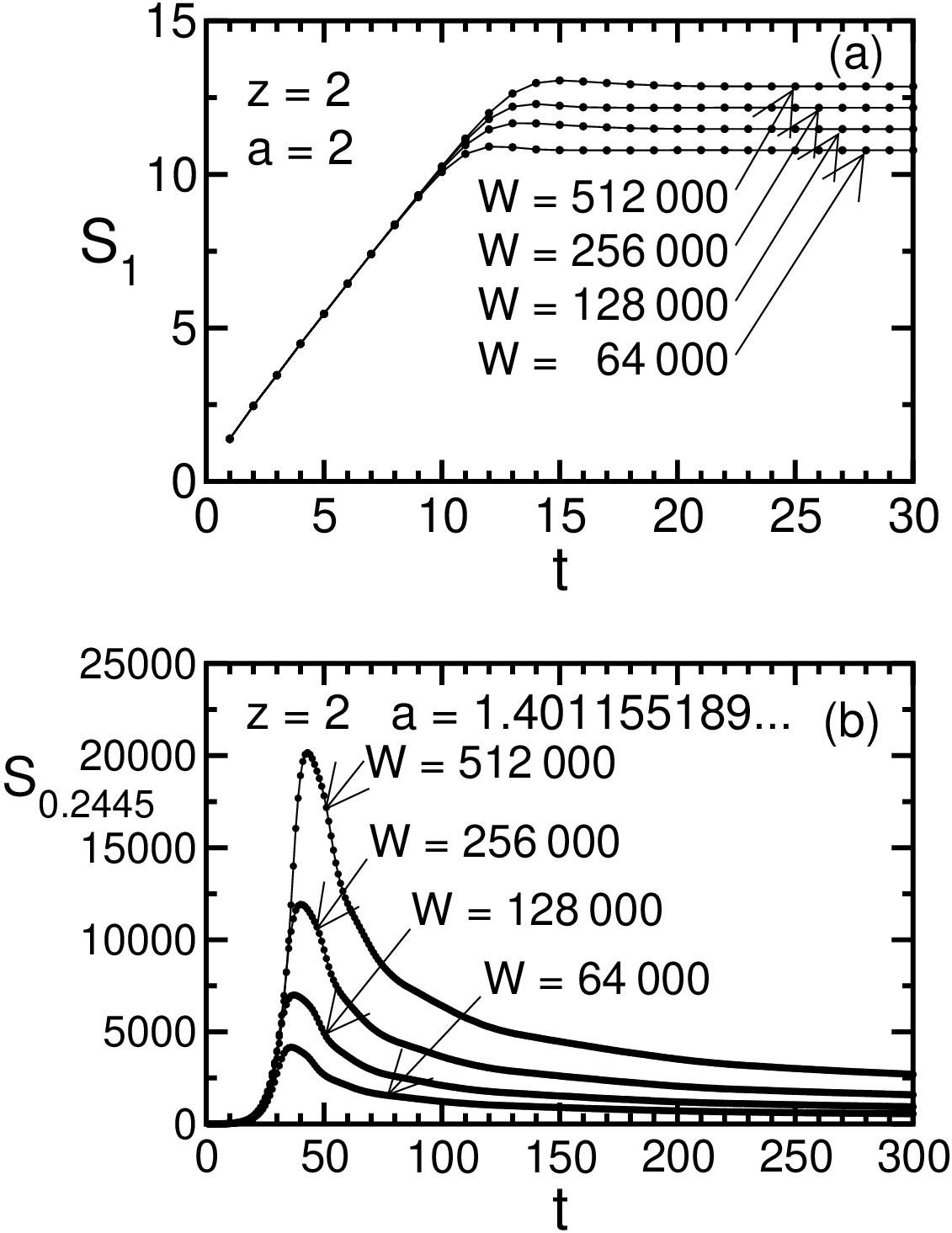}
\end{center}
\caption{\small
$S_{q_{sen}}(t)$ for $z=2$ . 
(a) Chaotic region, with $q_{sen}=1$; (b) Edge of chaos, with 
$q_{sen}=0.2445$, associated to  the region before the peak [12,14], and with $q_{rel}(W)>1$, associated to the region after the peak. 
(the highest value attained by $S_{0.2445}$, for a given $W$, is about 
$70 \%$ of the value corresponding to equiprobability).
Notice that for $a=2$, the stationary state is achieved for $t$ 
{\it much smaller} than that for $a=a_c$, which of course reflects the fact 
that relaxation is exponentially quick in the former (as it is also the case 
for $a<a_c$) whereas it is a power-law in the latter.
}
\label{Fig_1}
\end{figure}

\newpage
\begin{figure}[htb]
\begin{center}
\includegraphics[width=0.75\textwidth,keepaspectratio,clip=]{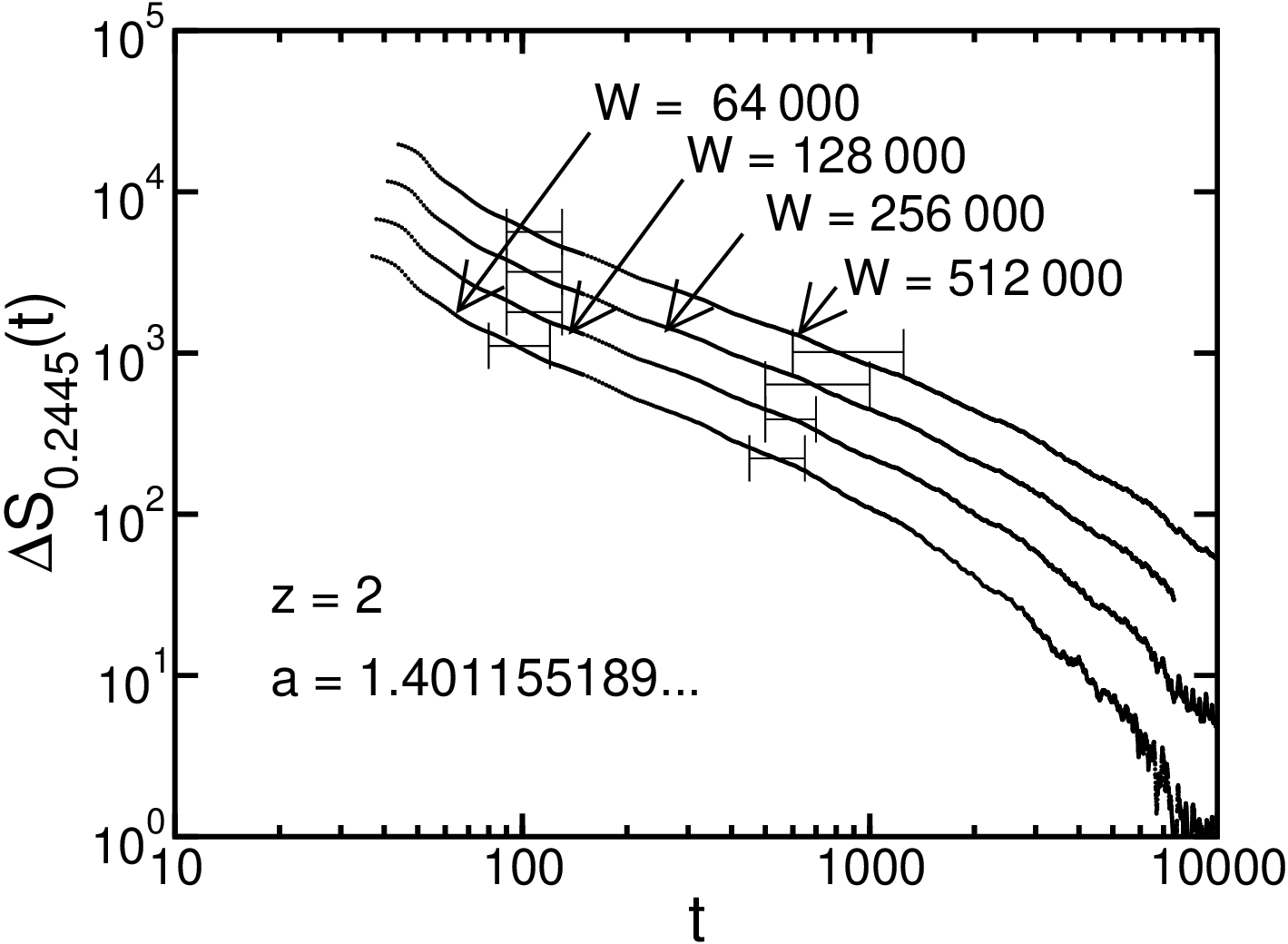}
\end{center}
\caption{\small
$\Delta S_{q_{sen}}(t) \equiv S_{q_{sen}}(t) - S_{q_{sen}}(\infty)$ at the $z=2$
chaos threshold, for typical $W$. For each $W$ we consider as the power-law region (linear correlation coefficient above 0.99) that going from the left  
interval to the right one. Furthermore, the values for $q_{rel}(W)$ in Fig. 3 were obtained by averaging over many starting (ending) points inside the left (right) interval.}
\label{Fig_2}
\end{figure}

\newpage
\begin{figure}[htb]
\begin{center}
\includegraphics[width=0.75\textwidth,keepaspectratio,clip=]{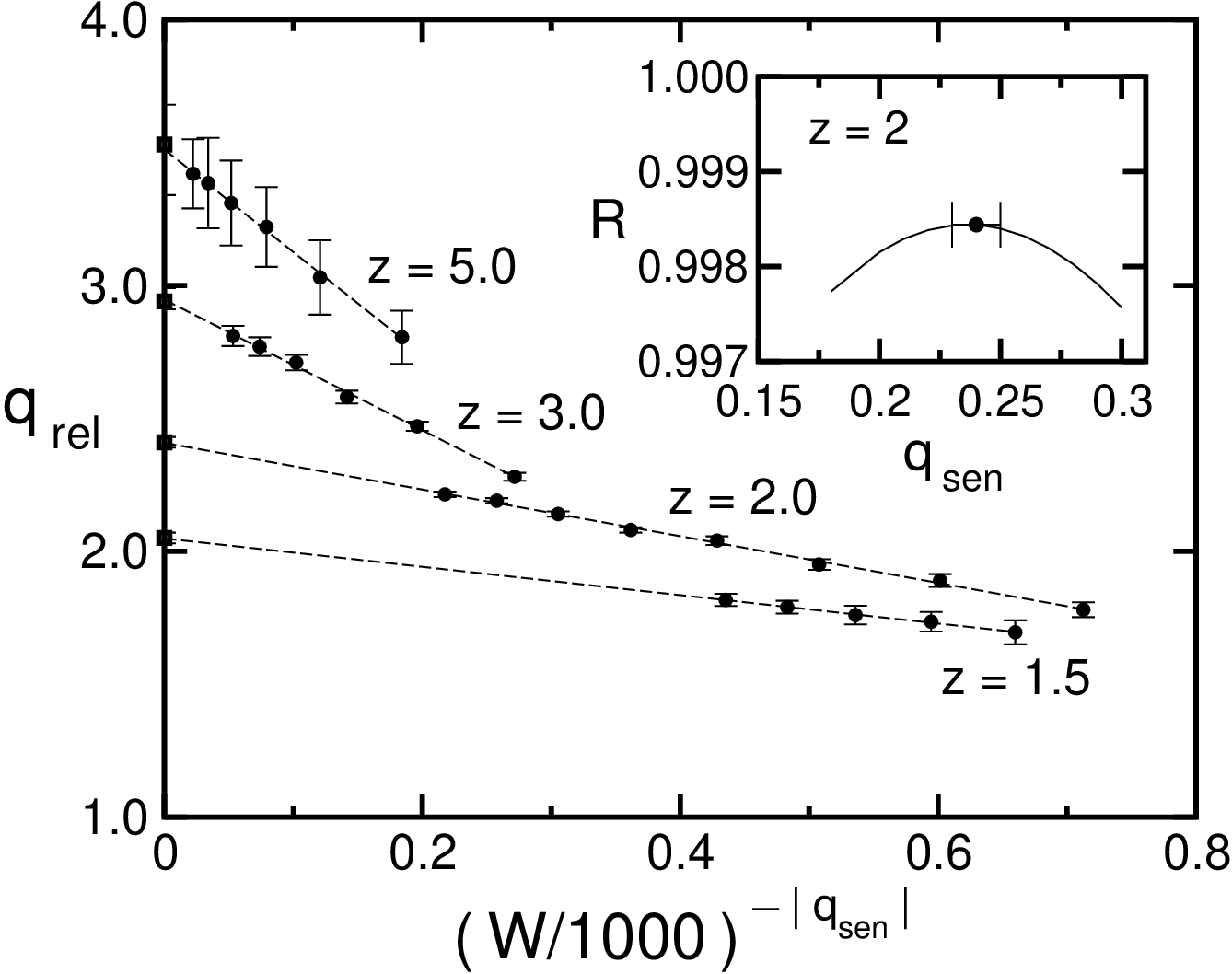}
\end{center}
\caption{\small 
$W$-dependence of $q_{rel}$ for typical values of $z$.
The $W \to \infty$ extrapolated values  coincide with 
the values reported in 
\protect\cite{Moura-Tirnakli-Lyra}. The abscissa has been chosen to be $(W/1000)^{-|q_{sen}|}$ (instead of $W^{-|q_{sen}|}$) for better visualization. Inset: Linear correlation coefficient procedure which has been used to select a specific numerical value for the sensitivity entropic index $q_{sen}(z)$.}
\label{Fig_3}
\end{figure}

\newpage
\begin{figure}[htb]
\begin{center}
\includegraphics[width=0.75\textwidth,keepaspectratio,clip=]{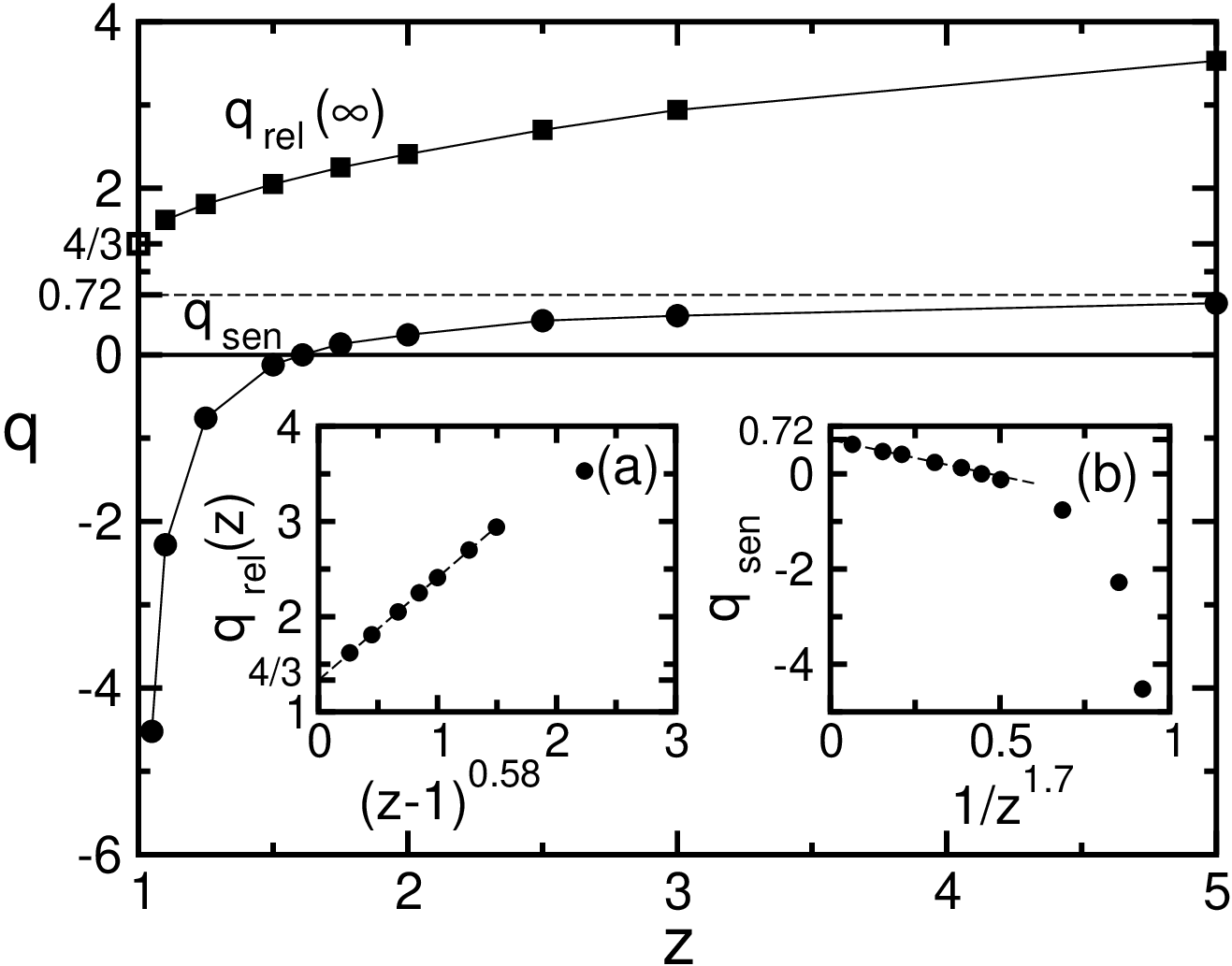}
\end{center}
\caption{\small
$z$-dependences of $q_{sen}$ (from \protect\cite{Costa-Lyra-Plastino-CT}, 
or self-consistently through Eq.~(\protect\ref{scaling}))
and $q_{rel}(\infty)$.
Inset (a): $z \rightarrow 1$ extrapolation for $q_{rel}(\infty)$;
$q_{rel}(\infty) = 4/3 + 1.077(z-1)^{0.58}$ fits the $z \le 3.0$ data.
Inset (b): $z \rightarrow \infty$ extrapolation for $q_{sen}(\infty)$;
$q_{sen}(z) = 0.72 - 1.525/z^{1.7}$ fits the $z \ge 1.75$ data.
}
\label{Fig_4}
\end{figure}

\end{document}